\RequirePackage[2020-02-02]{latexrelease}
\documentclass[aps,prc,twocolumn,showpacs,showkeys,nofootinbib,superscriptaddress]{revtex4-2}

\usepackage{longtable}
\usepackage{graphicx}
\usepackage{amsmath}
\usepackage{amsfonts}
\usepackage{amssymb}
\usepackage{natbib} % Use the natbib bibliography and citation package
\usepackage{setspace}
\usepackage{xspace}
\usepackage{cancel}
\usepackage{physics}
\usepackage{tensor,braket,color,bbold,dsfont,slashed}
\usepackage{stackengine}%Figure title stacking engine for LateX
\usepackage{newtxtext,newtxmath}
\usepackage{mathrsfs}

\usepackage[normalem]{ulem}

\renewcommand{\sout}{\bgroup \color{red} \ULdepth=-.5ex \ULset}

\begin{document}
\title{
%  \begin{flushright}
%    \rightline{PKNU-NuHaTh-2021-08} 
%  \end{flushright}
Gluon and valence quark distributions for the pion and kaon in nuclear matter}

\author{Parada T.P. Hutauruk}
\email{phutauruk@gmail.com}
\affiliation{Department of Physics, Pukyong National University (PKNU), Busan 48513, Korea}

\author{Seung-il Nam}
\email{sinam@pknu.ac.kr}
\affiliation{Department of Physics, Pukyong National University (PKNU), Busan 48513, Korea}
\affiliation{Center for Extreme Nuclear Matters (CENuM), Korea University, Seoul 02841, Korea}

\date{\today}

\begin{abstract}
In this paper we study the gluon and valence quark distributions in the pion and kaon in nuclear medium for various nuclear densities as well as in vacuum within the Nambu--Jona-Lasinio (NJL) model with the help of the proper-time regularization scheme which simulates a confinement of QCD. The nuclear medium effect is also determined in the same model for the symmetric nuclear matter. We then analyze the gluon and valence quark distributions for the bound pion and kaon in symmetric nuclear matter as well as those in vacuum. We find that the valence quark and gluon distributions in vacuum have relatively good agreements with the experimental data, the lattice QCD simulations, and the JAM Monte-Carlo (MC) global fit QCD analysis. Evolving to the higher factorization scale $Q = $ 4 GeV, the in-medium gluon and valence-quark distributions of the pion for various nuclear densities are turned out to be almost unchanged in comparison to the vacuum cases. On the contrary, for the kaon, they increase significantly with respect to the densities. Finally, we find that the vacuum gluon distribution for the kaon is smaller than that for the pion, which is consistent with other theoretical predictions. This feature holds for the in-medium gluon distribution in the nuclear density up to the saturation density.         
\end{abstract}

%\pacs{13.15.+g, 13.40.Gp, 25.30.Pt, 97.60.Jd, 14.60.Pq}

%\keywords{valence-quark distribution, gluon distribution, Nambu-Jona-Lasinio model, DGLAP QCD evolution, nuclear matter}

\maketitle

%========================================
\section{Introduction} \label{intro}
%=========================================

Parton distribution functions (PDFs) play important roles in describing the nonperturbative aspects of quantum chromoDynamics (QCD) for the internal structure of the hadron bound states~\cite{Berger:1979du}. Also, they are very crucial quantities for computing the cross sections for the high-energy hadron-hadron and neutrino-hadron interactions. Recently the valence quark and gluon distributions of the pseudoscalar (PS) mesons, i.e., kaon and pion become a more attractive subject in the hadronic physics community~\cite{Xie:2021ypc,Chavez:2021llq,Costa:2021mpk,Barry:2018ort,Arrington:2021biu,Nam:2012vm,Chen:2016sno,Hutauruk:2018zfk,Ding:2019lwe}. This is due to the fact that they can provide us a better understanding for the PS mesons'internal structures and the dynamics of quarks and gluons. In addition to these theoretical reasons, it is also triggered by the future experiments, i.e, the Electron-Ion Collider (EIC)~\cite{Arrington:2021biu,Aguilar:2019teb} and Electron-Ion Collider in China (EicC)~\cite{Anderle:2021wcy} as well as the COMPASS++/AMBER new QCD facility at CERN SPS~\cite{Adams:2018pwt}, which will be expected to provide more precise data with a wide range of the kinematic coverage for the gluon and quark distributions for the mesons.

It is widely known that pions and kaons are the Nambu-Goldstone bosons (NGBs) that emerged as a consequence of dynamical chiral symmetry breaking (D$\chi$SB). A deeper understanding of the internal structure of the mesons leads us to more profound insights of the D$\chi$SB and \textit{ vice versa}. Recently, remarkable progress has been achieved in studying the gluon distributions of the PS mesons in vacuum. Several theoretical models~\cite{Costa:2021mpk,Chen:2016sno,Ding:2019lwe,Lan:2021wok,Lan:2019vui}, the Jefferson Lab Angular Momentum Collaboration (JAM) phenomenology global fit QCD analysis~\cite{Barry:2018ort}, and lattice QCD simulations~\cite{Fan:2021bcr,Novikov:2020snp} have been applied to study the valence-quark and gluon distribution functions for the mesons in vacuum. However, besides these efforts, a theoretical understanding of those quark and gluon distributions still requires more studies, and the situation is worsened by the scarcity of experimental data that brings us to a difficulty in resolving the current controversies on the valence-quark distributions at high-$x$, for example, the power law of the distribution functions (DFs) at $x\rightarrow 1$~\cite{Holt:2010vj,Ball:2016spl} and the gluon distributions at low $x$~\cite{Ball:2016spl}, respectively. Attempts on studying the valence quark and gluon distribution functions in nuclear medium are much more limited.

Recent studies on the in-medium modifications of pion and kaon structures have been reported in the literature~\cite{deMelo:2014gea,Hutauruk:2019ipp} for investigating the pion valence-quark distribution amplitude (VDA) in medium using the light-front (LF) model, associated with the quark-meson coupling (QMC) model ~\cite{deMelo:2014gea} and the in-medium valence-quark distribution functions (VDFs) of the pion and kaon in the Nambu--Jona-Lasinio (NJL) model augmented with the QMC model~\cite{Hutauruk:2019ipp}. Similarly, but for the nucleon case, a very recent study was made to investigate the gluon distributions of nucleon in vacuum compared with those in the nuclei using the NJL model~\cite{Wang:2021elw}. They found the significant effects of the medium modifications for the unpolarized and polarized gluon distributions of bound nucleon in nuclear matter. Also, they reported that, in their work, the gluon distributions of the nucleon at scale $Q^2$ were dynamically generated via next-to-leading order (NLO) Dokshitzer-Gribov-Lipatov-Altarelli-Parisi (DGLAP) QCD evolution~\cite{Miyama:1995bd}, meaning that it is no gluon dynamics at initial scale.

Inspired by these recent studies, for the first time we study in the present work the pion and kaon gluon distributions in nuclear medium in the framework of the NJL model with the help of the proper-time regularization (PTR) scheme, simulating a confinement of QCD. Aside from focusing on the pion and kaon gluon distributions in nuclear medium, we also compute the pion and kaon valence-quark distributions in nuclear medium as well. The NJL model has been successfully applied for various physics phenomena of low-energy nonperturbative QCD, namely, the transverse momentum dependent (TMD)~\cite{Ninomiya:2017ggn}, the fragmentation function (FF)~\cite{Matevosyan:2013aka}, the pion and kaon properties in nuclear medium~\cite{Hutauruk:2018qku,Hutauruk:2019was} and the properties of neutron star~\cite{Tanimoto:2019tsl}.

In our approach, the nuclear medium effect is also calculated in the NJL model, which is the same model as used in the vacuum one. Thus, we first present our results for the pion and kaon gluon distributions in vacuum compared with the lattice QCD simulations~\cite{Fan:2021bcr,Novikov:2020snp} and the JAM phenomenology global fit QCD analysis~\cite{Barry:2018ort}. Note that the gluon distributions for the pion and kaon are absent at initial model scale as in Ref.~\cite{Wang:2021elw}, since there is no gluon dynamics in the NJL model. It is absorbed into the $G_\pi$ coupling constant. Hence, in this work, we purely generate the gluon and sea-quark distributions from the NLO DGLAP QCD evolution through the parton splitting functions. Next, we compute the gluon and valence quark distributions for the mesons in nuclear medium for various baryon densities to observe how the in-medium modifications change the distributions. This study is very helpful to shed light on the quark-gluon dynamics in the bound pion and kaon in symmetric nuclear matter (SNM).

This paper is organized as follows. In Sec.~\ref{sec:NJL} we briefly introduce the effective Lagrangian for the SU(3) flavor NJL model and the quark properties, namely, the constituent quark mass and meson-quark coupling constant that required in the calculation of the vacuum parton distribution functions. We then present an expression for the twist-2 valence quark distributions of the pion and kaon in vacuum. Moreover, we describe the SNM of the NJL model (SNM-NJL) that is used for computing the valence quark distributions in SNM. Finally, we present our formula for the in-medium distributions in the NJL model. In Sec.~\ref{results} our numerical results are presented and their implications are discussed. Section~\ref{summary} is devoted for a summary.
%

%====================================================
\section{Nambu--Jona-Lasinio Model}
\label{sec:NJL}
%=====================================================
%
%

In this section we briefly present the NJL effective Lagrangian and the properties of the pion and kaon in the model. It maintains the important features of the nonperturbative QCD, i.e. the spontaneously chiral symmetry breaking (S$\chi$SB) for instance. The SU(3) flavor NJL Lagrangian is given by
\begin{eqnarray}
  \label{eq1gpk}
  \mathscr{L}_{\textrm{NJL}} &=& \bar{q} \left( i \partial \!\!\!/ - \hat{m}_q \right) q + G_\pi [ (\bar{q} \mathbf{\lambda}_a q)^2 -(\bar{q} \mathbf{\lambda}_a \gamma_5 q )^2 ] \nonumber \\
  &-& G_\rho [(\bar{q} \mathbf{\lambda}_a \gamma^\mu)^2 + (\bar{q} \mathbf{\lambda}_a \gamma^\mu \gamma_5 q)^2].
\end{eqnarray}
With the quark fields $q$ are defined by $q = (u, d, s)^T$, $\hat{m}_q=\textrm{diag}(m_u, m_d, m_s)$ represents the current-quark mass matrix, and $\mathbf{\lambda}_a$ are the Gell-Mann matrices in flavor space with $\lambda_0 \equiv \sqrt{\frac{2}{3}} \mathds{1}$. The $G_\pi$ and $G_\rho$ are the coupling constants of the four-fermion dimensional with units of GeV$^{-2}$. Thus, the standard solution to NJL gap equation is given by
\begin{eqnarray}
  \label{eq2gpk}
  S_q^{-1} (p) &=& p\!\!\!/ - M_q + i \epsilon, 
\end{eqnarray}
where the subscript $q = (u, d, s)$ denotes the quark flavor and the dynamical quark mass $M_q$ in the PTR scheme is given by
\begin{eqnarray}
  \label{eq3gpk}
  M_q &=& m_q + \frac{3G_\pi M_q}{\pi^2} \int_{\tau_{\textrm{UV}}^2}^{\tau_{\textrm{IR}}^2} \frac{d\tau}{\tau^2} \exp \left( -\tau M_q^2 \right),
\end{eqnarray}
where $\tau_{\textrm{IR}}^2 = 1/\Lambda_{\textrm{IR}}^2$ and $\tau_{\textrm{UV}}^2 = 1 /\Lambda_{\textrm{UV}}^2$ stand for respectively the infrared (IR) and ultraviolet (UV) integration limits with the value of $\Lambda_{\textrm{IR}} =$ 0.240 GeV, which determined based on the limit of $\Lambda_{\textrm{QCD}}$ ($\simeq$ 0.2-0.3 GeV), and $\Lambda_{\textrm{UV}}$ are the infrared and ultraviolet cutoffs, respectively.

In the NJL model, pions and kaons, as bound state of the dressed quark-antiquark, can be obtained by solving the Bethe-Salpeter equations (BSEs). The BSE solutions are given by the interaction channel of the two-body amplitude. For those PS mesons, it has the form
\begin{eqnarray}
  \label{eq4gpk}
  t_{\pi, K} &=& \frac{-2i G_\pi}{1 + 2 G_\pi \Pi_{\pi, K} (p^2)},
\end{eqnarray}
where the polarization insertions for the PS mesons are respectively given by
\begin{eqnarray}
  \label{eq5gpk}
  \Pi_{\pi} (p^2) &=& 6i \int \frac{d^4 k}{ (2 \pi)^4} [\gamma_5 S_l (k) \gamma_5 S_l (k+p)], \\
  \label{eq5gpkb}
  \Pi_{K} (p^2) &=& 6i \int \frac{d^4 k}{ (2 \pi)^4} [\gamma_5 S_l (k) \gamma_5 S_s (k+p)].
\end{eqnarray}
Here, the subscripts of $l = (u,d$) and $s$ are respectively referring to the light and strange quarks. From the pole of the amplitude $t_{\pi,K}$ in Eq.~(\ref{eq4gpk}) we determine the PS-meson masses by solving the pole equations $1 + 2 G_\pi \Pi_\pi (p^2 = m_\pi^2)=$ 0 for the pion and $1 + 2 G_\pi \Pi_K (p^2 = m_K^2)=$ 0 for the kaon. By solving these equations analytically, the expressions for the PS-meson masses in the PTR scheme are straightforwardly obtained by
\begin{eqnarray}
  \label{eq6gpk}
  m_\pi^2 &=& \frac{m_l}{M_l} \frac{2}{G_\pi \mathscr{I}_{l l} (m_\pi^2)},  \\
   \label{eq6gpkb}
  m_K^2 &=& \left( \frac{m_s}{M_s} + \frac{m_l}{M_l} \frac{1}{G_\pi \mathscr{I}_{l s} (m_K^2)} + (M_s - M_l)^2 \right),
\end{eqnarray}
where the quantity of $\mathscr{I}_{ls} (p^2)$ in Eq.~(\ref{eq6gpkb}) is defined for the kaon by
\begin{eqnarray}
  \label{eq7gpk}
  \mathscr{I}_{ls} (p^2) &=& \frac{3}{\pi^2} \int_0^1 dz \int_{\tau_{\textrm{UV}^2}}^{\tau_{\textrm{IR}^2}} \frac{d\tau}{\tau} \nonumber \\
  &\times& \exp \left( -\tau (z (z-1) p^2 + z M_s^2 + (1-z) M_l^2) \right).
\end{eqnarray}
For the pion case, one of the quark flavors is changed as $s\to l$, giving $\mathscr{I}_{l l} (p^2) $.

Thus, straightforwardly, the meson-quark coupling constants can be determined through the residue at pole in the quark-antiquark amplitude $t_{\pi,K}$. In other word, it can be simply obtained by calculating the first derivative of the polarization insertion in Eqs.~(\ref{eq5gpk}), (\ref{eq5gpkb}) for the corresponding PS meson with respect to the $p^2$. It has a form
\begin{eqnarray}
  \label{eq8gpk}
  g_{m q\bar{q}} ^{-2} &=& -\frac{\partial \Pi_m (p^2)}{\partial p^2} \Big|_{p^2 = m_m^2}, 
\end{eqnarray}
with the subscript $m = (\pi, K)$. The meson-quark coupling constant of the $g_{m q\bar{q}}$ has a relation with the wave function renormalization constant that gives $Z_m = g_{m q\bar{q}} ^{-2}$.

%====================================================
\subsection{Vacuum pion and kaon parton distributions}
\label{sec:vacpdf}
%=====================================================
%

In this section we present the formulas for the valence quark distributions of the PS mesons in vacuum. The twist-2 quark distributions are simply defined by
\begin{eqnarray}
  \label{eq9gpk}
  q_{m} (x) &=& \frac{p^{+}}{2\pi} \int d\xi^{-} \exp \left( ix p^{+} \xi^{-} \right) \langle m| \bar{q}(0) \gamma^{+} q (\xi^{-})| m \rangle_{c}, \nonumber \\
\end{eqnarray}
where $x = \frac{k^{+}}{p^{+}}$ is the Bjorken scaling variable or the longitudinal momentum of the parton in the PS mesons with $k^{+}$ is the plus-component of the struck momentum of quark and $p^{+}$ is the plus-component of the PS meson momentum, $\xi$ is the skewness variable, and the subscript $c$ denotes the connected matrix element. 
\begin{figure}[b]
\centering\includegraphics[width=\columnwidth]{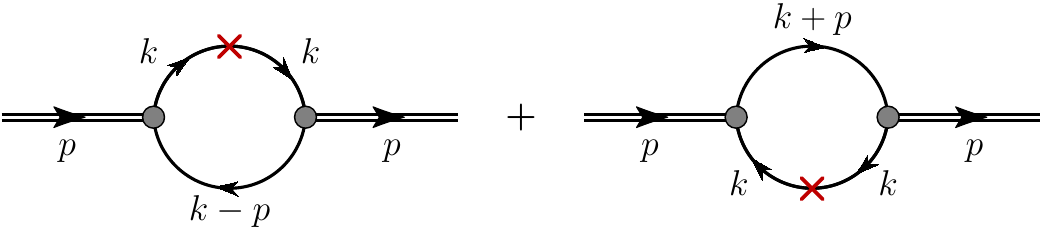}
\caption{\label{fig1a} (Color online) The relevant diagrams for the PS-mesons valence quark distributions. The red crossed is an operator insertion 
$\gamma^+ \delta \left( p^+x - k^+ \right) \hat{P}_q$, where $\hat{P}_q$ is the projection operator for quarks of flavor $q$.}
\end{figure}
Following our previous work in Ref.~\cite{Hutauruk:2016sug}, we then calculate the valence quark distributions based on two Feynman diagrams in Fig.~\ref{fig1a}. The expressions for the operator insertion of the light and strange quarks are given as follows
\begin{eqnarray}
  \label{eq10gpk2}
\gamma^+ \delta \left( k^{+} - x p^{+} \right) \hat{P}_{l} &=& \gamma^+ \delta \left( k^{+} - x p^{+} \right) \frac{1}{2} \left( \frac{2}{3} \mathds{1} \pm \lambda_3 + \frac{1}{\sqrt{3}} \lambda_8 \right),
\cr
\gamma^+ \delta \left( k^{+} - x p^{+} \right) \hat{P}_{s} &=& \gamma^+ \delta \left( k^{+} - x p^{+} \right) \left( \frac{1}{3} \mathds{1} - \frac{1}{\sqrt{3}} \lambda_8 \right).
\end{eqnarray}
Considering the relation $\bar{q} (x) = - q(-x)$, the valence-quark distributions can be defined by
\begin{eqnarray}
  \label{eq10gpk}
  q_{\pi,K} (x) &=& i g_{m q \bar{q}}^2 \int \frac{d^4k}{(2\pi)^4} \delta \left( k^{+} - xp^{+} \right) \nonumber \\
  &\times& \mathrm{Tr}_{c,f,\gamma} [\gamma_5 \lambda_a^{\dagger} S_l (k) \gamma^{+} \hat{P}_{u/d} S_l (k) \gamma_5 \lambda_a S_l (k-p)], \nonumber \\
  \bar{q}_{\pi,K} (x) &=& -i g_{m q \bar{q}}^2 \int \frac{d^4k}{(2\pi)^4} \delta \left( k^{+} + xp^{+} \right) \nonumber \\
  &\times& \mathrm{Tr}_{c,f,\gamma} [\gamma_5 \lambda_a S_l (k) \gamma^{+} \hat{P}_{\bar{d}/\bar{s}} S_l (k) \gamma_5 \lambda_a^{\dagger} S_s (k+p)], \nonumber \\
\end{eqnarray}
where the meson-quark coupling constant of the $g_{m q \bar{q}}$ that defined in Eq.~(\ref{eq8gpk}). The trace runs over the color, flavor, and Lorentz indices. Similar with Ref.~\cite{Hutauruk:2016sug}, these valence quark distributions are evaluated using the moment, which is defined by
\begin{eqnarray}
  \label{eq11gpk}
  \mathscr{A}_n &=& \int_0^1 dx x^{(n-1)} q_{\pi,K} (x),
\end{eqnarray}
with $n$ is an integer number. We then apply the Ward-Takahashi-like identity $S(k) \gamma^+ S(k) = - \partial S(k) / \partial k_{+}$ and perform the Feynman parametrization.

For general interest, here, we present the explicit expression for the valence quark distributions of the kaon in the PTR scheme:
\begin{eqnarray}
  \label{eq12gpk}
  q_{K} (x) &=& \frac{3 g_{K q \bar{q}}^2}{4\pi^2} \int_0^1 dx \int^{\tau_{IR}^2}_{\tau_{UV}^2} d\tau \nonumber \\
  &\times& \exp \left( -\tau (x(x-1)m_K^2 + xM_s^2 + (1-x) M_l^2) \right) \nonumber \\
  &\times& \Big[ \frac{1}{\tau} + x (1-x) \left( m_K^2 - (M_l -M_s)^2 \right)\Big], \nonumber \\
\end{eqnarray}
and for the valence antiquark distribution of the kaon is given by
\begin{eqnarray}
  \label{eq12gpkb}
  \bar{q}_{K} (x) &=& \frac{3 g_{K q \bar{q}}^2}{4\pi^2}  \int_0^1 dx \int^{\tau_{IR}^2}_{\tau_{UV}^2} d\tau \nonumber \\
  &\times& \exp \left( -\tau (x(x-1)m_K^2 + xM_l^2 + (1-x) M_s^2) \right) \nonumber \\
  &\times& \Big[ \frac{1}{\tau} + x (1-x) \left( m_K^2 - (M_l -M_s)^2 \right)\Big].
\end{eqnarray}
Thus, that of the pion is straightforwardly obtained by replacing $M_s \rightarrow M_l$ and $g_{K q \bar{q}}^2 \rightarrow g_{\pi q \bar{q}}^2$ in Eqs.~(\ref{eq12gpk}), (\ref{eq12gpkb}), resulting the $u_\pi (x) = \bar{d}_\pi (x)$.

The valence quark distributions must satisfy the baryon number and momentum sum rules. The expression for the baryon number and momentum sum rules for the PS mesons are respectively given by
\begin{eqnarray}
  \label{eq13gpk}
  \int_0^1 dx [u_K (x) - \bar{u}_K (x)] = \int_0^1 dx [\bar{s}_K -s_K (x)] &=& 1, \nonumber \\
  \int_0^1 dx\,x[u_K (x) + \bar{u}_K (x) + s_K (x) + \bar{s}_K (x)] &=& 1. 
\end{eqnarray}
Here we again emphasize that, in our NJL model, the PS-meson sea-quark and gluon distributions are zero at initial scale $Q_0^2$, since in the NJL model the gluons are absorbed into the $G_\pi$ coupling constant. Hence the NJL model model has no dynamical gluons as in Ref.~\cite{Wang:2021elw}. So, they are purely generated using the NLO DGLAP QCD evolution in present work.

In the DGLAP QCD evolution, the parton distributions for the PS meson are classified into three types; the gluon distribution, the quark distribution, and the sea-quark distribution. The quark distribution, which is well known as the valence quark distribution, is the nonsinglet (NS) quark distribution, which is simply defined by
\begin{eqnarray}
  q_\mathrm{NS} (x) &=& q(x) - \bar{q} (x),
\end{eqnarray}
where $q (x)$ and $\bar{q}(x)$ are respectively the quark and antiquark distributions. The evolution of the nonsinglet quark distribution in the DGLAP QCD evolution are defined by
\begin{eqnarray}
  \frac{\partial q_{\textrm{NS}} (x,Q^2)}{\partial \ln (Q^2)} &=& \mathscr{P}_{qq} \left(x,\alpha_{\textrm{s}}(Q^2)\right) \otimes q_{\textrm{NS}} (x,Q^2),
\end{eqnarray}
where $\mathscr{P}_{qq}$ denotes the splitting of the q-q (quark-quark) function. The physics interpretation of $\mathscr{P}_{qq}$ is the probability for a quark of type $q$ with momentum fraction $z$ to emit the quark and becomes a quark of new type $q$ with momentum fraction $x$. A convolution product between the splitting function and the nonsinglet quark distribution has the form
\begin{eqnarray}
\mathscr{P}_{qq} \otimes q_{\mathrm{NS}} &=& \int_x^1 \frac{dz}{x} \mathscr{P} \left(\frac{x}{z} \right) q_{\mathrm{NS}}(z,Q^2).
\end{eqnarray}

The other type of quark distribution is called a singlet quark distribution and it is defined by
\begin{eqnarray}
q_\mathrm{S} (x) = \sum_i q_i^+ = \sum_i q_i (x) + \bar{q}_i (x),
\end{eqnarray}
where $i$ is the quark flavor. The singlet quark distributions are expressed by
\begin{eqnarray}
  \label{eqdglap}
  \frac{\partial}{\partial \ln Q^2} \begin{bmatrix} q_\mathrm{S} (x,Q^2) \\ g(x,Q^2) \end{bmatrix} &=& \begin{bmatrix} \mathscr{P}_{qq} & \mathscr{P}_{qg} \\ \mathscr{P}_{gq} & \mathscr{P}_{gg} \end{bmatrix} \otimes \begin{bmatrix} q_\mathrm{S} (x,Q^2) \\ g(x,Q^2) \end{bmatrix}.
\end{eqnarray}
Hence, the gluon distributions for the PS mesons can be obtained by solving Eq.~(\ref{eqdglap}) numerically. In analogy to Taylor expansion series, the splitting functions can be also expanded in terms of the running coupling constant, $\alpha_{s} (Q^2)$ in perturbative region and it then takes a form
\begin{eqnarray}
  \mathscr{P} (z,Q^2) &=& \left(\frac{\alpha}{2\pi} \right) \mathscr{P}^{(0)} (z) + \left(\frac{\alpha}{2\pi} \right)^2 \mathscr{P}^{(1)} (z) + \cdot \cdot \cdot,
\end{eqnarray}
where the first term in $\mathscr{P}^{(0)} (z)$ is the leading order (LO), the second term in $\mathscr{P}^{(1)} (z)$ denotes the NLO. The NLO result is expressed by
\begin{eqnarray}
  \alpha_{s} &=& \frac{4\pi}{\beta_0} \frac{1}{\ln (Q_{\Lambda})} \left[ 1 - \frac{\beta_1}{\beta_0} \frac{\ln \ln (Q_{\Lambda})}{\ln (Q_{\Lambda})}\right] + \mathscr{O} \left( \frac{1}{\ln^2 (Q_{\Lambda})}\right),
\end{eqnarray}
with
\begin{eqnarray}
  Q_{\Lambda} &=& \frac{Q^2}{\Lambda_{QCD}^2},~~~~~\beta_0 = \frac{11}{3}N_c - \frac{4}{3}N_f, \nonumber \\
  \beta_1 &=& \frac{34}{3} N_c^2 - \frac{10}{3} N_c N_f - 2 C_F N_f.
\end{eqnarray}
Here $N_c$, $N_f$ are the number of colors and the number of active flavors, respectively and $C_F = \frac{4}{3}$. The value of $\Lambda_{QCD}$ depends on the number of active flavors and the renormalization scheme.

%=====================================================
\subsection{Nuclear matter NJL model}
\label{sec:nuclearmatter}
%=====================================================
%
%

Here we present the equation of state (EoS) for SNM of the NJL model. Further details can be found in Ref.~\cite{Bentz:2001vc} and references therein. Using the quark bilinears that defined by $\bar{q} \Gamma q = \langle \rho| \bar{q} \Gamma |\rho \rangle + : \bar{q} \Gamma q :$ with the vertices $\Gamma = \mathds{1}, \gamma^\mu$, the NJL effective Lagrangian in Eq.~(\ref{eq1gpk}) is modified in SNM as follows: 
\begin{eqnarray}
  \label{eq13agpk}
\mathscr{L}_{\mathrm{SNM-NJL}} &=& \bar{q} (i \partial \!\!\!/ - M_q - V\!\!\!\!/) q - \frac{(M_q - m_q)^2}{4G_\pi} + \frac{V_\mu V^\mu}{2G_\omega} \nonumber \\
  &+& \mathscr{L}_{I} , 
\end{eqnarray}
where the in-medium constituent quark mass and the isoscalar-vector mean field are defined respectively
\begin{eqnarray}
  \label{eq13bgpk}
  M_q &=& m_q - 2 G_\pi \langle \rho| \bar{q} q |\rho \rangle, \\
   \label{eq13bgpkb}
  V^\mu &=& 2 G_\omega  \langle \rho| \bar{q} \gamma^\mu q |\rho \rangle = 2 \delta^{0 \mu} G_\omega \langle q^{\dagger} q \rangle ,
\end{eqnarray}
and $\mathscr{L}_I$ is the interaction Lagrangian. The $G_\pi$ and $G_\omega$ are the scalar and vector coupling constants. The vector potential is defined as $V =(V_0,\mathbf{0})$.

Using the hadronization technique, the effective potential for SNM can be obtained for the NJL Lagrangian. In the mean-field approach, the effective potential is simply given by
\begin{eqnarray}
  \label{eq14gpk}
  \mathscr{E} &=& \mathscr{E}_V - \frac{V_0^2}{4 G_\omega} + 4 \int \frac{d^3 p}{(2\pi)^3} \theta \left( p_F - |\mathbf{p}| \right) \epsilon_p ,
\end{eqnarray}
where $\epsilon_p = \sqrt{M_N^{*2} + \mathbf{p}^2} + 3V_0 \equiv E_p + 3 V_0$. Note that the effective nucleon mass as a function of the constituent quark $M_N^* = M_N^* (M_q^*)$ determined by solving the quark-diquark bound state for the nucleon using the relativistic Faddeev equation. The vacuum contribution of the quark loop in Eq.~(\ref{eq14gpk}) is given by
\begin{eqnarray}
  \label{eq15gpk}
  \mathscr{E}_V &=& 12 i \int \frac{d^4 k}{(2\pi)^4} \ln \left( \frac{k^2 -M^2 + i\epsilon}{k^2 - M_0^2 + i \epsilon} \right) + \frac{(M - m)^2}{4 G_\pi} \nonumber \\
  &-& \frac{(M_0 - m)^2}{4 G_\pi}.
\end{eqnarray}
In the present study, we consider the SNM at rest. Using the condition $\partial \mathscr{E} / \partial V_0 = 0$, we then determine the value of $V_0$ as
\begin{eqnarray}
  \label{eq16gpk}
  V_0 &=& 6 G_\omega \rho_B,
\end{eqnarray}
where $\rho_B = 2 p_F^3 / 3 \pi^2$ is the baryon density with $p_F$ is the Fermi momentum for the baryon. Also, the constituent quark mass $M_q$ for fixed baryon density must satisfy the condition $\frac{\partial \mathscr{E}}{\partial M_q} = 0 $ to give the similar expression of the in-medium gap equation, as in Eq.~(\ref{eq13bgpk}).

\begin{figure*}[t]
  \centering\stackinset{r}{5cm}{t}{0.5cm}{(a)}{\includegraphics[width=0.45\textwidth]{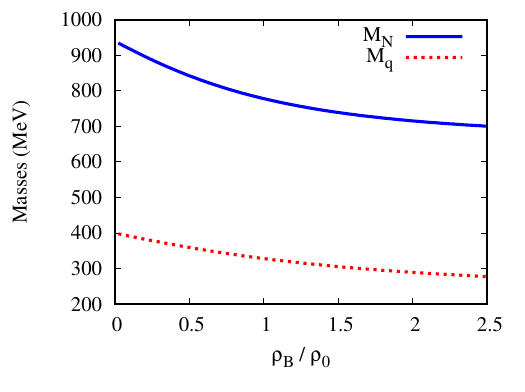}}\hspace{0.2 cm}
  \centering\stackinset{r}{5cm}{t}{0.5cm}{(b)}{\includegraphics[width=0.45\textwidth]{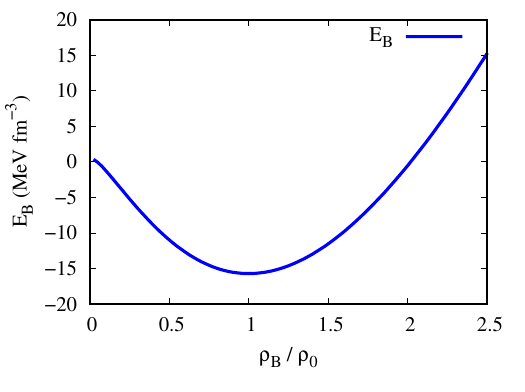}}
  \caption{\label{fig1b} (a) The effective nucleon and constituent quark masses as functions of the baryon density ($\rho_B/\rho_0$). (b) The binding energy per nucleon as a function of the baryon density.}
\end{figure*}

A result for the effective nucleon and in-medium constituent quark masses as a function of baryon density $\rho_B / \rho_0$ is shown in Fig.~\ref{fig1b} (a). It shows that the effective nucleon and in-medium quark constituent masses decrease as the baryon density increases, as we expected, indicating the partial restoration of the spontaneous breakdown of chiral symmetry in medium.

Thus the expression for the binding energy per nucleon is given by
\begin{eqnarray}
  \label{eq17gpk}
  \frac{E_B}{A} &=& \frac{\mathscr{E}}{\rho_B} - M_{N0}.  
\end{eqnarray}
With the $M_{N0}$ is the nucleon mass in vacuum. A numerical result for binding energy as a function of baryon density $\rho_B / \rho_0$ is depicted in Fig.~\ref{fig1b} (b). It shows that the SNM-NJL model can reproduce well the binding energy per nucleon $E_B/A =$ -15.7 MeV at saturation density $\rho_0 =$ 0.16 fm$^{-3}$. The details of the obtained parameters in the SNM-NJL model will be described in Sec.~\ref{results}.

%==================================================================
\subsection{In-medium pion and kaon parton distributions}
\label{sec:medpdf}
%===================================================================
%
%
In this section the valence quark distributions of the PS mesons in SNM are presented. Following the vacuum case, for general example, we present the quark and antiquark distributions for the $K^+$-meson, which contains [$u\bar s$] quarks, as
\begin{eqnarray}
  \label{eq18gpk}
  u_{K^+}^{} (x)  &=& \frac{3\,g_{K q q}^{*2}}{4\pi^2} \int_0^1 dx \int^{\tau_{IR}^2}_{\tau_{UV}^2} d\tau\,
  \nonumber \\ && \mbox{}
   \times \left[\frac{1}{\tau} + x(1 - x)\left[m_K^{* 2} - (M_l^{*} 
       - M_s^{*})^2\right] \right]
   \nonumber \\ && \mbox{}
  \times e^{-\tau \left[ x(x - 1)\,m_K^{* 2} + x\,M_s^{* 2} + (1-x)\,M_l^{* 2} \right]}, \\ 
  \label{eq19gpk}
  \bar{s}_{K^+}^{}(x)  &=& \frac{3\,g_{K q q}^{*2}}{4\pi^2} \int_0^1 dx \int^{\tau_{IR}^2}_{\tau_{UV}^2} d\tau\
  \nonumber \\ && \mbox{}
  \times \left[\frac{1}{\tau} + x(1 - x)
    \left[m_K^{* 2} - (M_l^{*} - M_s^{*})^2\right]\right]
   \nonumber \\ && \mbox{}
   \times  e^{-\tau \left[ x(x - 1)\,m_K^{* 2} + x\,M_l^{* 2} +  (1-x)\,M_s^{* 2} \right]} .
\end{eqnarray}
The valence quark distributions of the $\pi^+$ in SNM can be easily obtained by replacing $M_s^{*} \to M_l^{*}$ and $g_{K q q}^{*} \to g_{\pi q q}^{*}$, which gives the symmetry relation of $u_{\pi^+}(x) = \bar{d}_{\pi^+}(x)$.

The valence quark distributions of the PS mesons in SNM and vacuum with the corresponding Bjorken variables $\tilde{x}_{a}$ and $x_a$, respectively, are related by~\cite{Hutauruk:2019ipp}
\begin{align}
  \label{eq:valence6a}
  q_{K^{+}}^{} (x_a) &= \frac{\epsilon_F^{}}{E_F} q_{K^{+}}^{*} ( \tilde{x}_a )
\end{align}
with
\begin{equation}
  \tilde{x}_a = \frac{\epsilon_F^{}}{E_F} x_a - \frac{V^0}{E_F},
\end{equation}
where the quark energy in SNM $\epsilon_F = \sqrt{k_F^2 + M_q^{*2}} + V^0 \equiv E_F + V^0$. The $k_F$ is the quark Fermi momentum which is related to the quark density as $\rho_q^{} = 2k_F^3/\pi^2$. For the antiquark, it is simply given by $\epsilon_F^{} = E_F - V^0$. The above formulas are valid only for light ($u$, $d$) quarks in the present approach. Note that the in-medium valence quark distributions satisfy the baryon number and momentum sum rules.

%====================================================
\section{Numerical result} \label{results}
%===================================================
Our numerical results for the in-medium and vacuum gluon and valence quark distributions for the pion and kaon compared with the lattice QCD simulations~\cite{Fan:2021bcr,Novikov:2020snp} and the JAM phenomenology global fit QCD analysis~\cite{Barry:2018ort} are presented. The parameters of the present NJL model determined in vacuum are the coupling constants $G_\pi$, $G_\omega$, $G_\rho$, $G_a$, $G_s$, and $\Lambda_{\textrm{UV}}$ as in Refs.~\cite{Tanimoto:2019tsl,Hutauruk:2016sug}. In the PTR scheme, we fix the regularization parameter of $\Lambda_{\textrm{IR}} =$ 0.240 GeV, which is in order of the $\Lambda_{\textrm{QCD}}$, and opt the dressed light constituent quark mass $M_l =$ 0.4 GeV. The remaining parameters are then fit to the physical masses of the pion $m_\pi =$ 0.14 GeV, kaon $m_K =$ 0.495 GeV, nucleon mass $M_N = M_{N0} =$ 0.94 GeV, and $\rho$-meson $m_\rho =$ 0.77 GeV along with the decay constant of the pion $f_\pi =$ 0.093 GeV. This choice gives $\Lambda_{\textrm{UV}} =$ 0.645 GeV, $G_\pi =$ 19.04 GeV$^{-2}$, $G_a =$ 2.8 GeV$^{-2}$, $G_s =$ 7.49 GeV$^{-2}$, and $M_{qs} =$ 0.611 GeV, where $G_s$, $G_a$, and $M_{\textrm{qs}}$ are respectively the scalar diquark coupling constant, the axial-vector diquark coupling constant, and the strange constituent quark mass. Note for the $\phi$ mass we get $m_\phi =$ 1.001 GeV, and the $G_s$ and $G_a$ coupling constants are determined in the Faddeev equations to fit the vacuum nucleon mass $M_{N}$ and axial coupling constant of nucleon $g_A =$ 1.267~\cite{Tanimoto:2019tsl}. Thus the scalar $M_s$ and axial $M_a$ diquark masses are calculated using these model parameters and we obtain $M_s =$ 0.687 GeV and $M_a =$ 1.027 GeV. The SNM parameters are determined to fit the binding energy per nucleon  $E_B /A =$ -15.7 MeV at saturation density $\rho_0 =$ 0.16 fm$^{-3}$, we then obtain $G_\omega = $ 6.03 GeV$^{-2}$.

%
%
%%%  FIG 1
\begin{figure*}[t]
  \centering\stackinset{r}{5cm}{t}{0.5cm}{(a)}{\includegraphics[width=0.45\textwidth]{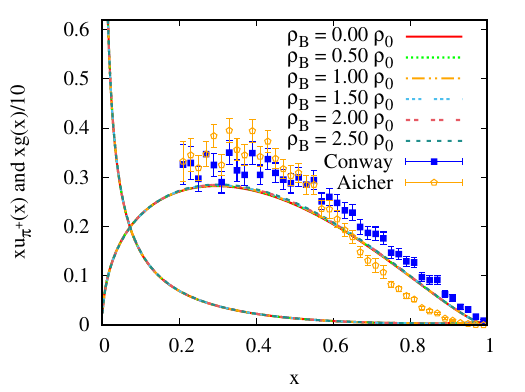}}
  \centering\stackinset{r}{5cm}{t}{0.5cm}{(b)}{\includegraphics[width=0.45\textwidth]{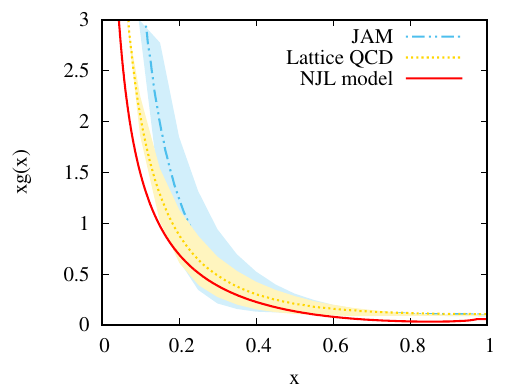}}
  \caption{\label{fig1} (a) Valence quark and gluon distributions of the pion in SNM for various nuclear densities as well as in vacuum as functions of the longitudinal momentum $x$ and (b) the pion gluon distribution in vacuum compared with the lattice QCD simulation~\cite{Fan:2021bcr} and JAM global fit analysis~\cite{Barry:2018ort}. All DFs are evolved from the initial model scale of $Q_0^2 =$ 0.16 GeV$^2$ to a scale $Q^2 =$ 16 GeV$^2$ using the NLO DGLAP QCD evolution. The experimental data are the E615-Conway (blue shaded region)~\cite{Conway:1989fs} and reanalysis E615 (orange shaded region)~\cite{Aicher:2010cb}.}
  \centering\stackinset{r}{5cm}{t}{0.5cm}{(a)}{\includegraphics[width=0.45\textwidth]{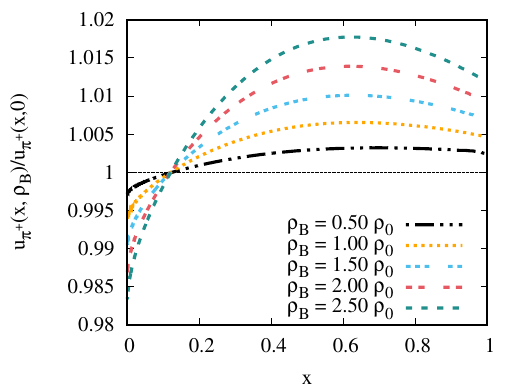}}\hspace{0.2 cm}
  \centering\stackinset{r}{5cm}{t}{0.5cm}{(b)}{\includegraphics[width=0.45\textwidth]{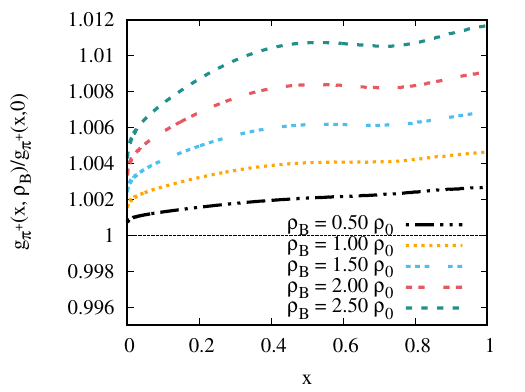}}
  \caption{\label{fig2} The ratios of (a) the valence quark distributions of the pion in SNM for various nuclear densities to the pion vacuum valence-quark distributions and (b) for those of the pion gluon distributions.}
\end{figure*}
%%%%%%%%%%%%%%%%%%
%
%
Results for the in-medium valence quark and gluon distributions of the pion as a function of the light-front momentum fraction $x$ for various baryon densities as well as that in vacuum that are evolved from the initial model scale $Q_0^2 =$ 0.16 GeV$^2$~\cite{Hutauruk:2016sug} to a scale at $Q^2=$ 16 GeV$^2$ using the NLO DGLAP evolution shown in Fig.~\ref{fig1} (a). It indicates that the valence quark distribution for the pion seems less significant change in nuclear medium. This result is consistent with the obtained result in Ref.~\cite{Hutauruk:2019ipp}. The behaviors of the in-medium valence quark PDFs are followed by the pion gluon distributions in nuclear medium. In order to verify our approach for the gluon PDFs, a comparison of our result for the pion gluon distribution in vacuum with the lattice QCD simulation~\cite{Fan:2021bcr}, and the JAM phenomenology global fit QCD analysis~\cite{Barry:2018ort} is presented in Fig.~\ref{fig1} (b). Our result shows a good agreement with the results from the lattice QCD simulation (orange dotted line)~\cite{Fan:2021bcr}, and the JAM phenomenology global fit QCD analysis (blue dashed-dotted line)~\cite{Barry:2018ort}.

%%%  FIG 3
\begin{figure*}[t]
  \centering\stackinset{r}{5cm}{t}{0.5cm}{(a)}{\includegraphics[width=0.45\textwidth]{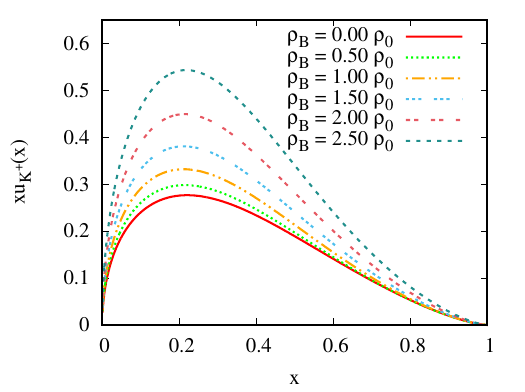}} \hspace{0.2 cm}
  \centering\stackinset{r}{5cm}{t}{0.5cm}{(b)}{\includegraphics[width=0.45\textwidth]{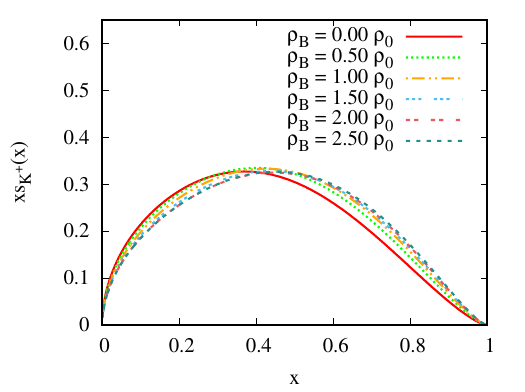}}\\
  \centering\stackinset{r}{5cm}{t}{0.5cm}{(c)}{\includegraphics[width=0.45\textwidth]{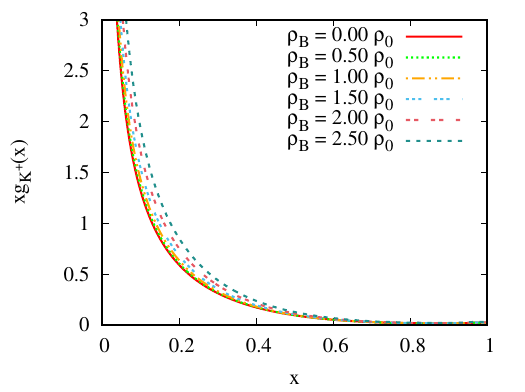}}
  \caption{\label{fig3} (a) Up valence-quark distributions of the kaon, (b) the strange valence-quark distributions of the kaon, and (c) the gluon distributions of the kaon.}
\end{figure*}
%%%%%%%%%%%%%%%%%%
%
%

The ratios of the pion valence-quark distributions in SNM to those in vacuum for various densities are shown in Fig.~\ref{fig2} (a). Figure~\ref{fig2} (a) indicates that the ratios decrease as the nuclear density increases up to $x \simeq 0.2$, which is consistent with the finding in Ref.~\cite{Chang:2021utv}. However, the reduction is relatively small for $x \lesssim 0.2$. This is because, in the region of $x \lesssim 0.2$, the pion gluon distribution is expected relatively large and it becomes growing in nuclear medium that might be suppressed the valence quark DFs in the pion. In contrast, for $x \gtrsim$ 0.2, the ratios increase with respect to the nuclear matter density. The ratios of the gluon distributions for the pion are depicted in Fig.~\ref{fig2} (b) in the same manner. It shows that the ratios of the pion gluon distributions in nuclear medium for various densities to those in vacuum increase relatively weakly with respect to the density.

\begin{figure*}[t]
  \centering\stackinset{r}{5cm}{t}{0.5cm}{(a)}{\includegraphics[width=0.45\textwidth]{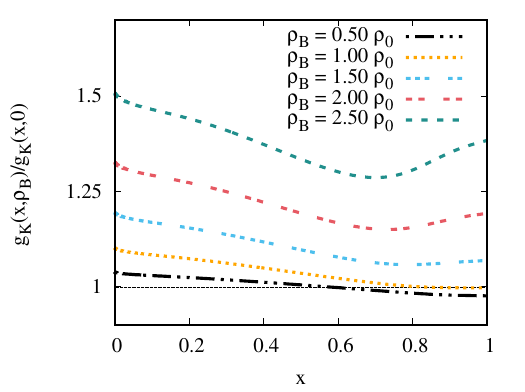}} \hspace{0.2 cm}
  \centering\stackinset{r}{5cm}{t}{0.5cm}{(b)}{\includegraphics[width=0.45\textwidth]{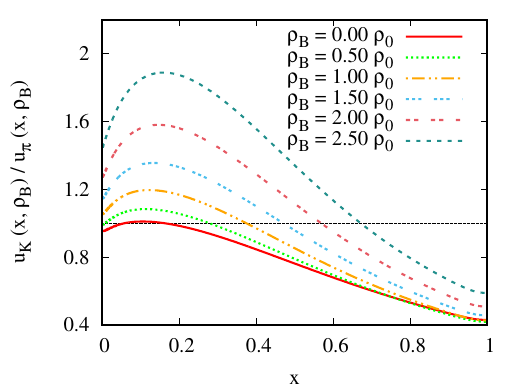}}\\
  \centering\stackinset{r}{5cm}{t}{0.5cm}{(c)}{\includegraphics[width=0.45\textwidth]{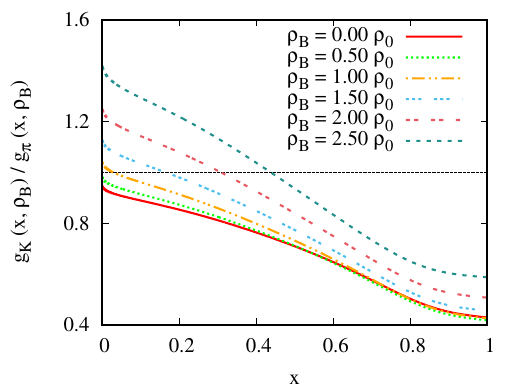} }\hspace{0.2 cm}
  \centering\stackinset{r}{5cm}{t}{0.5cm}{(d)}{\includegraphics[width=0.45\textwidth]{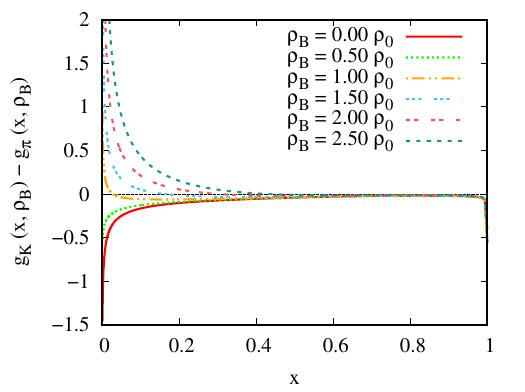}}  
  \caption{\label{fig4} The ratios for (a) the gluon distributions of the kaon in SNM to the vacuum gluon distributions, (b) the up valence-quark distributions of the kaon to the up valence-quark distributions of the pion, (c) the gluon distributions of the kaon to the gluon distributions of the pions and (d) the gluon distribution differences for the kaon and pions in SNM for various nuclear densities.}
\end{figure*}

For the kaon case, the kaon valence-quark distributions are evolved from the initial model scale $Q_0^2 =$ 0.16 GeV$^2$ to $Q^2=$ 16 GeV$^2$ using the NLO DGLAP evolution and depicted in Fig.~\ref{fig3} (a). We find that the up valence-quark distributions for the kaon increase with respect to the density. The enhancement of the up valence-quark distributions for the kaon in nuclear medium are relatively significant. It is more pronounced at higher densities. The results for the strange quark distributions in nuclear medium for various densities as well as those in vacuum after evolved at $Q^2 =$ 16 GeV$^2$ are shown in Fig.~\ref{fig3} (b). We find that the strange valence-quark distributions of the kaon decrease as the density increases in the range $0 \lesssim x \lesssim 0.4$. Thus, at $x \gtrsim 0.4$, the strange valence-quark distributions of the kaon begin to increase as the density increases. The gluon distributions of the kaon in nuclear medium are shown in Fig.~\ref{fig3} (c). In contrast to the gluon contents of the pion in nuclear medium for various densities, it seems that the gluon distributions of the kaon in nuclear medium increase more significantly as the density increases. The enhancement on the gluon distributions for the kaon in nuclear medium can be clearly seen in Fig.~\ref{fig4} (a). Also, the ratios of the up valence-quark distributions of the kaon to those of the pion in nuclear medium are given in Fig.~\ref{fig4} (b). It shows that the ratios of the up valence-quark distributions of the kaon to those of the pion in nuclear medium increase as the density increases and it decreases as the $x$ increases.
%
%
%%%  FIG 4
%%%%%%%%%%%%%%%%%%
%
%
%

The ratios of the gluon distributions of the kaon to those of the pion as a function of $x$ for various nuclear densities are shown in Fig.~\ref{fig4} (c). It can be seen that the ratios of the gluon distributions in the kaon increase with respect to the densities. However, it decreases as the $x$ increases. It implies the gluon distributions in the pion are relatively larger than those in the kaon. This can be understood because the strange quark, which has heavier mass than the light quark, radiates less instantly than the light quark and it radiates soft gluons than the light quark does. This finding is compatible with the result obtained in the Dyson-Schwinger equations (DSEs) model~\cite{Chen:2016sno}, which also pointed out in Ref.~\cite{Ding:2019lwe}. Figure~\ref{fig4} (c) depicts that the larger gluon distribution contribution in the pion occurs at $\rho_B \lesssim \rho_0$, which gives the ratios of $g_K (x,\rho_B) / g_\pi (x,\rho_B) \lesssim$ 1. Unexpectedly, we find that the ratios of $g_K (x,\rho_B) / g_\pi (x,\rho_B) \gtrsim$ 1 for $\rho_B > \rho_0$. This indicates that the gluon distribution in the kaon becomes larger than those in the pion for the higher density region ($\rho_B > \rho_0$).

To more clearly understand this unexpected gluon contribution, we calculate the gluon distribution differences for the pion and kaon as shown in Fig.~\ref{fig4} (d). We find that Fig.~\ref{fig4} (d) supports what we found in Fig.~\ref{fig4} (c). At $\rho_B \lesssim \rho_0$, it gives $g_K (x,\rho_B) - g_\pi (x,\rho_B) < $ 0 (negative value), but for $\rho_B > \rho_0$, we find that $g_K (x,\rho_B) - g_\pi (x,\rho_B) > $ 0 (positive value). The negative values of the $g_K (x,\rho_B) - g_\pi (x,\rho_B)$ clearly explains that the gluon content in the pion is significantly larger than that in the kaon in vacuum and the nuclear density up to the saturation density, $\rho_B \simeq \rho_0$. The positive values mean that the gluon content in the pion is relatively lower than that in the kaon in vacuum and the nuclear medium at $\rho_B > \rho_0$.

\begin{figure}[t]
  \centering{\includegraphics[width=0.45\textwidth]{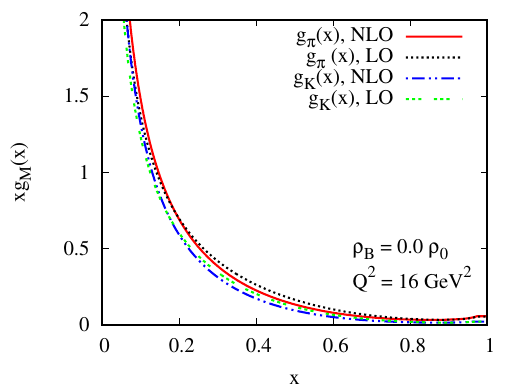}}
  \caption{\label{fig5} Vacuum gluon distributions for the pion and kaon via the LO and LO$+$NLO calculations at $Q^2 =$ 16 GeV$^2$.}
\end{figure}

Finally, in order to test the DGLAP evolution dependencies, we present the numerical result for the vacuum gluon distributions of the pion and kaon via the LO and NLO calculations at $Q^2 =16$ GeV$^2$ in Fig.~\ref{fig5}. It is clearly shown that, for both pion and kaon, the NLO contributions provide qualitatively small change in comparison to those from the LO ones. We also verified that the $Q^2$ and density dependencies of the NLO contributions are considerably weak.  

%
%
%%%  FIG 5
%%%%%%%%%%%%%%%%%%
%
%
%

%==================================
\section{Summary} \label{summary}
%=================================
%

To summarize, we have studied the in-medium and vacuum valence-quark and gluon distributions of the pion and kaon in the Nambu--Jona-Lasinio model with the help of the proper-time regularization scheme, simulating a QCD confinement. To describe nuclear matter, we adopt the SNM-NJL model. We then computed the valence quark distributions of the pion and kaon in nuclear medium for various nuclear densities as well as in vacuum. In this work, at initial scale $Q_0^2$, the gluon distributions in the pion and kaon are taken to be zero, since  there is no gluon dynamics in the NJL model. So, the gluon distributions in the pion and kaon in SNM for various nuclear densities as well as in vacuum are purely generated from the NLO DGLAP QCD evolution.

We have determined the gluon and quark distributions in SNM at a higher scale by evolving them from the initial scale $Q_0^2 =$ 0.16 GeV$^2$ to a factorization scale at $Q^2 =$ 16 GeV$^{2}$. Our result for the vacuum gluon and valence-quark distributions of the pion have good agreement with the results obtained from the JAM phenomenology global fit QCD analysis~\cite{Barry:2018ort}, the lattice QCD simulation~\cite{Fan:2021bcr}, and the experimental data~\cite{Conway:1989fs}, respectively. The results for the in-medium gluon and valence-quark distributions in the pion for various nuclear densities shows that their distributions are almost unchanged in nuclear medium.

The results for the kaon case, the gluon and up valence-quark distributions for various nuclear densities significantly modify in nuclear medium, in particular, for the in-medium up valence-quark distributions in the kaon. The in-medium strange valence-quark distributions in the kaon for various densities decrease as the nuclear densities increase at $x \lesssim 0.4$. In contrast, it becomes increasing as the nuclear density increases at $x \gtrsim 0.4$.

Our results clearly show that the vacuum gluon distributions in the kaon is smaller than those in the pion, which is consistent with the DSEs result in Ref.~\cite{Chen:2016sno}. Also, we found that the in-medium gluon distributions in the kaon is smaller than those in the pion up to $\rho_B \simeq \rho_0$. Surprisingly, for higher nuclear densities $\rho_B > \rho_0$, the gluon distributions in the kaon are found to be bigger than those in the pion as shown in Fig.~\ref{fig4} (d). This finding may be useful and relevant information for the compact star (neutron star) or heavy-ion collisions studies. In such cases, the NJL model serves as a very useful tool to guide possible future computations for the lattice QCD and other more sophisticated theoretical approaches.

\begin{acknowledgments}
P. T. P. H thanks X. G. Wang and Ian Clo\"et for useful discussions. This work was supported by the National Research Foundation of Korea (NRF) funded by the Korea government (MSIT) Grants No.~2018R1A5A1025563 and No.~2019R1A2C1005697.
\end{acknowledgments}

\end{document}